# Predicting digital asset market based on blockchain activity data

Zvezdin Besarabov
National School of Mathematics and Natural Sciences
Sofia, Bulgaria
me@zvezd.in

Todor Kolev
Comrade Cooperative
Sofia, Bulgaria
todor@comradecoop.com

## ABSTRACT

Blockchain technology shows significant results and huge potential for serving as an interweaving fabric that goes through every industry and market, allowing decentralized and secure value exchange, thus connecting our civilization like never before.

The standard approach for asset value predictions is based on market analysis with an LSTM neural network. Blockchain technologies, however, give us access to vast amounts of public data, such as the executed transactions and the account balance distribution. We explore whether analyzing this data with modern Deep Leaning techniques results in higher accuracies than the standard approach.

During a series of experiments on the Ethereum blockchain, we achieved 4 times error reduction with blockchain data than an LSTM approach with trade volume data. By utilizing blockchain account distribution histograms, spatial dataset modeling, and a Convolutional architecture, the error was reduced further by 26%.

The proposed methodologies are implemented in an open source cryptocurrency prediction framework, allowing them to be used in other analysis contexts.

## 1 INTRODUCTION

Cryptocurrencies, the decentralized new ways of exchanging value, are gaining a lot of interest recently with the emergence of Bitcoin, Ethereum, and hundreds of other cryptoassets with total value that recently exceeded $500 billion. Unlike the traditional markets, where all the data about the market and the trading itself are centralized and controlled by gatekeeper organizations, the cryptoasset markets are public, decentralized and transparent by definition.

The new kind of data available on the blockchain can be analyzed together with standard market information in an attempt to predict future market movements.

Our goal is to explore whether we can utilize the abundant blockchain data using modern Deep Learning techniques to estimate future facts about cryptoassets.

In this paper, we are introducing a set of methodologies for the cryptocurrency prediction problem and demonstrate their performance on the Ethereum (*ETH*) cryptocurrency (Wood 2018). First, raw blockchain and financial data is collected (Secton 3), which is used in different feature extraction algorithms (Section 4). The features are compiled into a dataset using different modeling techniques (Section 5). Custom neural architectures are trained, evaluated and compared on their predictive performance (Section 7 and 8).

### 1.1 Related work

Projects about cryptoasset market predictions have emerged in the past, including a paper by a Stanford University team (Madan 2014) and two more recent Github projects (Bynum 2015) (Remy 2017). All of them focus primarily on Bitcoin price estimations, do not explore the rich publicly available blockchain data in depth, and do not use or have not disclosed sophisticated deep learning techniques for their estimations.

## 2 BACKGROUND

The following subsections describe the background information for the research.

### 2.1 Blockchain and Ethereum

Satoshi Nakamoto's introduction of Bitcoin in November 2008 (Nakamoto 2009) has often been hailed as a radical development in money and currency as it's the first example of a digital asset which simultaneously has no backing or "intrinsic value" and no centralized issuer or controller. Another arguably more important part of the Bitcoin experiment is the underlying blockchain technology as a tool of distributed consensus. The most important aspect of such technology is the absence of an intermediary (centralized server, bank, company, etc.) between the originator and the recipient, as any changes to the data on this chain are made by consensus among all members of a decentralized network. Thus, avoiding the need to trust third parties. The blockchain can be thought of as a distributed public database with records for each transaction in history. All cryptoasset operations and activity are contained within the blockchain transaction data.

The Ethereum project (*ETH*) is a newer implementation (Wood 2018), based on the blockchain technology. In addition to the transaction record keeping functionality, Ethereum provides a mechanism for executing program logic on each transaction, thus enabling a much wider variety of use cases of the blockchain technology. Some of which include financial transactions, secure voting, autonomous organizations, company management, freedom of speech networks, online games, crowdfunding, speculation, making it one of the more noisy and difficult to predict blockchains.

### 2.2 Autonomous decentralized applications

The intent of Ethereum is to create an alternative protocol for building autonomous decentralized applications. Because of the blockchain consensus, no one has control over the code execution or storage of these applications. A decentralized application can not be modified outside of the constraints of its logic. Because such an immutable application resembles contractual relations, autonomous decentralized applications are often referred to as "crypto contracts", "smart contracts" or simply contracts.

## 2.3 Transactions and traces

Every block in the chain is a package of transactions. Every transaction denotes a transfer of value and/or an execution of a smart contract.

Transactions only contain information about the initial (outside) call. For example, if a contract receives funds from $x$ and forwards them to $y$, the transaction will only indicate that the contract received the funds. The transfers the contract additionally makes are known as internal calls. To gather information about them, the initial transaction needs to be replayed in order to observe the derivation from it. Transfers, as well as other contract activity collected using this method are called *traces*.

## 2.4 Factors for digital asset value change

In basic economics, the correlation between supply and demand (Whelan et al. 2001) determines the price of an asset. Higher demand or lower supply relates to higher price.

Because of small market capitalization, digital currencies are unstable and volatile. Based on our observations of Ethereum, we have identified multiple factors as having a significant influence on the supply and demand of the currency. Most of them can be traced by analyzing the blockchain data in depth.

*2.4.1 Speculative investment.* Speculative investment is the action of buying and holding a certain asset, speculating that its value will rise in the future. This hampers Ethereum's already low supply and results in artificial price increases.

*2.4.2 Initial Coin Offerings.* Intial Coin Offerings (ICOs) (Catalini and Gans 2017) are share resemblance to Initial Public Offerings, but are implemented using smart contracts and therefore do not bear any of the legal protections that regular investments provide.

During 2017, the total cumulative ICO funding has increased by more than 80 times (Coindesk 2017). Ethereum's smart contract framework has become the most targeted platform for ICO fundraising. This is believed to have been a major reason for Ethereum's rise during June 2017.

*2.4.3 Altcoin interference.* After Bitcoin, hundreds of other cryptocurrencies have emerged. Despite the fact that they are completely unrelated and independent from each other, major events in one cryptocurrency can cause unexpected fluctuations in others. For example, major ICO events in Ethereum have caused surges in numerous other currencies, like Litecoin (CoinGecko 2017).

*2.4.4 Media influence.* Crypto market's value can also be easily influenced by media publications. A prime example of this is a case of widespread hoax about an incident with Ethereum's founder Vitalik Buterin (Roberts 2017). This led to temporary plummet and loss of 4\$ billion from Ethereum's market capitalization.

Currently our prediction models do not take into account the media influence. Our goal is in a further research to perform deep-learning based sentiment analysis on news and social media sources related to cryptocurrencies.

## 2.5 Use of Deep Learning in asset value predictions

Deep Learning (LeCun et al. 2015) allows for the discovery of patterns in a large dataset. Such a set consists of called dataset samples (forming input-output pairs, such as summarized blockchain activity and the future price). The inputs pass through multiple layers (which perform data transformations with free parameters) and leave as an output. The free parameters have to be set (trained, learned) on known input-output pairs.

Long Short-Term Memory (LSTM) (Hochreiter and Schmidhuber 1997) and Convolutional (CNN) (Lecun et al. 1998) networks are types of DL networks, which are well known for their performance in time series and image analysis respectively.

*2.5.1 Normalization.* Every value in a dataset needs to be normalized (scaled, fit), because of the numerical instability of large computer floating point operations. The general objective for a normalized series is to consist of small numbers of zero mean and equal variance. We use the following normalization algorithms depending on the situation:

$$basic_i = \frac{x_i - min(x)}{(max(x) - min(x))}$$

This basic min-max scale is used to map independent sequences with the same sign to the interval $\in [0; 1]$.

$$around\_zero_i = \frac{x_i + max(|max(x)|, |min(x)|)}{2 * max(|max(x)|, |min(x)|)}$$

Similar to *basic*, but maps positive and negative inputs to $(0.5; 1]$ and $[0; 0.5)$ respectively. This scale is used with sequences of varying sign.

$$image_i = (x_i - \frac{1}{n}\sum_{t=1}^{n} x_t) * \frac{1}{std(x)}$$

An an algorithm which produces unbounded time series of zero mean and equal variance. It is used with image-line sequences.

*2.5.2 Error measures.* The following measurement factors are often used to evaluate the performance of predictive DL:

$$sign = \frac{t}{t+f}; \quad err(t) = out_t - true_t; \quad null(t) = 0.5 - true_t$$

$$MSE = \frac{1}{n}\sum_{t=1}^{n} err(t)^2; \; RMSE = \sqrt{\frac{1}{n}\sum_{t=1}^{n} err(t)^2}; \; R^2 = 1 - \frac{\sum_{t=1}^{n} err(t)^2}{\sum_{t=1}^{n} null(t)^2}$$

Where *out* and *true* are lists of outputs and expected outputs respectively. If the outputs are binary, then the amount of correctly and incorrectly predicted flags is $t$ and $f$ respectively.

The metrics $R^2$ and *sign* measure accuracy $\in [0, 1]$, this the aim is to maximize them. In $R^2$ the performance of the model is compared to that of a model always returning the same value. *MSE* and *RMSE* are measures of error which have to be minimized.

Traditionally, *MSE* is used as the loss function during training. The rest are used for performance analysis.

## 3 RAW DATA

The first kind of data gathered is historical market tick data, which is aggregated from multiple exchanges to achieve a less biased view of the financial state. The frequency (size of each tick) is one hour. Every tick consists of the *open*, *close*, *low*, and *high* prices, as well as the trade volume to and from the currency for that period.

The second kind of data is from the Ethereum blockchain. It grows every day (Etherscan 2018) and includes hundreds of gigabytes worth of transactions, cryptocontract executions, and blockchain events for every moment its existence. For each block in the chain, the following data points are extracted: creation timestamp, number (chain index), miner (block creator), list of confirmed transactions, size in bytes, creation difficulty, and computational resources used (Gas limit and Gas used). The following is stored for each transaction in a block: address of the initiator and receiver, transferred value, used resource units (Gas), and amount paid per resource unit (Gas price). The same kind of information is also collected for internal contract activity by calculating transaction traces.

The total size of the gathered raw blockchain features is around 500GB, containing 5, 300, 000 blocks with a total of 194, 000, 000 transactions and close to a billion traces. It took 30 consecutive days to download and another 14 to process, filter, format, and save the data to a database.

Both the market and blockchain data are collected for the interval from 8-08-2015 to present.

## 4 GENERATION OF DATA PROPERTIES

In order to create a blockchain-based dataset, we first need to extract the most valuable information from the raw blockchain data by performing different feature extraction algorithms. A feature (also called property) is calculated once for each market tick in the historical data in order to form a time series.

We are extracting a set of properties defined in Table 1. They are created and selected based on their importance and significance to the prediction problem. A property name with suffix "_rel" denotes that its values have been converted to relative ones. An example for a property is seen in Figure 1.

The following subsections explain more advanced concepts in calculating properties.

### 4.1 Distributions

Other than extracting singular value properties like the discussed so far, we are also investigating ways to extract more valuable information from the blockchain by tracking the activity of the crypto accounts in higher detail. This is achieved by creating account distributions- 2D matrices that visualize account activity based on multiple account features. These distributions contain spatial value, which Convolutional networks can take advantage of.

Let us introduce some common definitions: $scl$ as an array of scale functions, $feat$ as an array of functions that return an account feature, and $mx$ as an array of constants.

Possible values for $scl_i$ include $\log_2$ and $\log_{1.2}$. Possible account features ($feat_i$) are defined in Table 2.

| Property | Description |
| --- | --- |
| openPrice | The value of ETH at the start of the tick |
| closePrice | The value of ETH at the end of the tick |
| stickPrice | *openPrice − closePrice* |
| volumeTo | Exchange volume to that cryptocurrency |
| volumeFrom from | Exchange volume from that cryptocurrency |
| transactionCount | Amount of transactions |
| dappOperations | Amount of transactions to crypto contracts |
| blockSize | The average size of a block in bytes |
| difficulty | The average difficulty for block mining |
| uniqueAccounts | Number of accounts in existence |
| gasLimit | Average limit on computational resource use |
| gasPrice | Average price per gas unit |
| gasUsed | Average amount of gas used |
| networkHashrate | Combined hashrate of every miner in the network |
| ETHSupply | Total amount of Ether in circulation |
| pendingTx | How many transactions per minute are pending for inclusion in the next block |
| blockchainGrowth | Growth of the blockchain size in gigabytes |

Table 1: List and description of the properties compiled from the raw data.

| Feature | Description |
| --- | --- |
| balance | Account balance, measured in *wei* ($10^{18}$ *wei* = 1 *ETH*). |
| lastSeen | Amount of seconds since the account's last participation in a transaction. |
| volumeIn | Amount of received value in the last tick, in *wei*. |
| volumeOut | Amount of sent value in the last tick, in *wei*. |
| transactionN | Amount of transactions where the account as either a receiver or a sender. |
| ERC20 | Amount of ERC20 token operations in the last tick (Ethereum contracts only) |

Table 2: The possible account feature functions.

### 4.2 Account balance distribution

This distribution visualizes summarized recent account activity in terms of the exchanged volume and amount of transactions, distributed based on balance groups.

Let us define $scl_0 = \log_2$ $mx_0 = 10^{26}$ $feat_0 = volumeIn$ $feat_1 = volumeOut$ $feat_2 = transactionN$ $groupN = \lfloor scl_0(mx_0) \rfloor$ $featN = 3$, $S$ as the accounts that have participated in a transaction in the current market tick, and *distribution* as a matrix of shape ($featN, groupN$).

For $acc \in S$:
$\quad gr = \min(\lfloor scl_0(balance(acc)) \rfloor, groupN - 1)$
$\quad$ For $x \in [0, featN)$:
$\quad\quad distribution_{x,gr} + = feat_x(acc)$
$distribution = scl_0(distribution)$

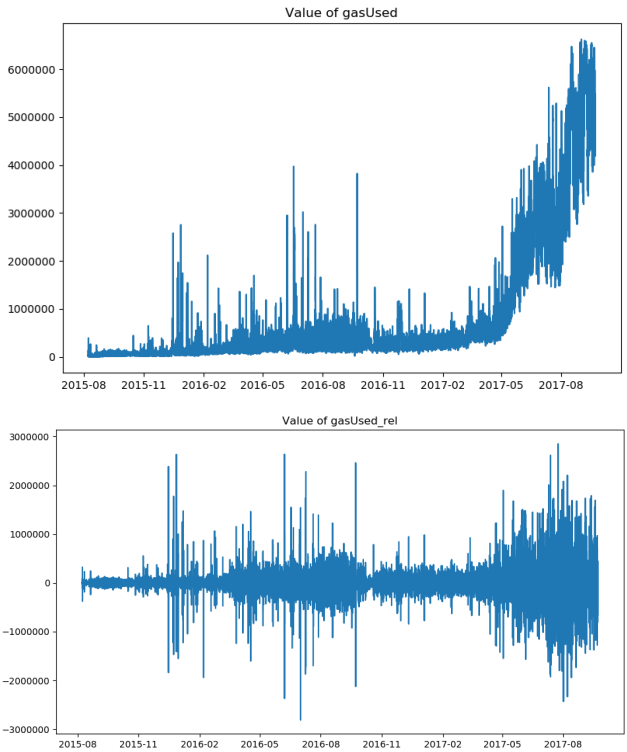

Figure 1: The absolute (up) and relative (down) values of property *gasUsed* over time. We can observe how major events have influenced the usage of the Ethereum network over time.

The final operation is to scale the large values of volumeIn and volumeOut.

Figure 2 contains an example of 2 such distributions.

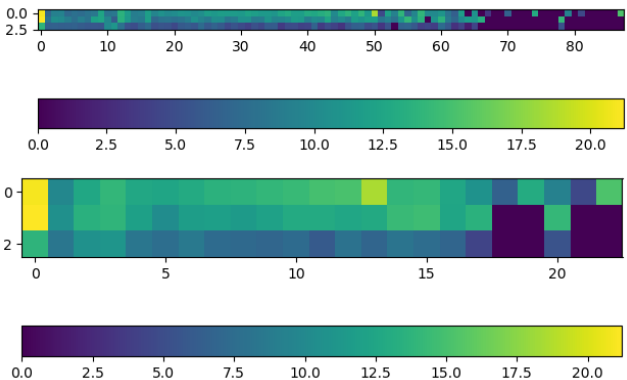

Figure 2: Examples with $scl_0 = \log_{1.2}$ (up) and $scl_0 = \log_2$ (down). We observe how the summarized account trade activity changes in the different balance groups.

### 4.3 Account number distributions

These multivariate distributions represent how the accounts are clustered based on 2 of their account features.

The process of creating an account number distribution is described as follows:

Repeat for $x = 1$ and $x = 2$:
$\quad groupN_x = \lfloor scl_x(mx_x) \rfloor$
For $acc \in S$, where $S$ is a chosen subset of (or all) accounts, do:
$\quad$Repeat for $x = 1$ and $x = 2$:
$\quad\quad group_x = \min(\lfloor scl_x(feat_x(acc)) \rfloor, groupN_x - 1)$
$\quad distribution[group_1, group_2] += 1$
Finally, for every value, do:
$\quad distribution = \log_2(distribution)$

The final $\log_2$ scaling is to mitigate the uneven distribution of accounts. The possible options for the *feat* functions are defined in Table 2. Some configurations of distributions are presented in Table 3 and are later referred to in the experiments.

| N | S | $feat_1$ | $feat_2$ | $scl_1$ | $scl_2$ | $mx_1$ | $mx_2$ |
|---|---|---|---|---|---|---|---|
| 1 | all | balance | lastSeen | $\log_{1.2}(\lfloor x/10^{17} \rfloor)$ | $\log_{1.2}$ | $10^7$ | $20736e3$ |
| 2 | contracts | balance | lastSeen | $\log_{1.2}(\lfloor x/10^{17} \rfloor)$ | $\log_{1.2}$ | $10^7$ | $20736e3$ |
| 3 | contracts | volumeIn | ERC20 | $\log_2(\lfloor x/10^{17} \rfloor)$ | $\log_2$ | $10^7$ | $262144$ |

Table 3: The configurations for account number distributions. The balances and transfers, measured in wei, are scaled to larger units ($10^{17}$ *wei* = 0.1 *eth*) to reduce noise.

Consequently, we refer to the distribution configurations as follows: balanceLastSeenDistribution (N=1), contractBalanceLastSeenDistribution (N=2), and contractVolumeInERC20Distribution (N=3). Example distributions are visualized in Figure 3.

## 5 GENERATION OF A DATASET

A dataset is a set of samples, containing inputs (normalized property data) and expected outputs (prediction target). The concepts and methodologies behind dataset generation are defined in the following subsections.

### 5.1 Prediction target

The prediction target is the future value a chosen target property. Predictions are for the duration of one market tick. Examples of a prediction target include relative price fluctuations, amount of new accounts, trade volume, transaction count and network clogging (a serious issue in the light of recent events (Baqer et al. 2015)).

Our experiments will be mainly focused on predicting price movements and new accounts.

### 5.2 Normalization

Every property in a dataset is a different time series and is normalized separately with a chosen algorithm (among those defined in Secion 2.5.1). A method to automatically determine the algorithm for a given property is called *prop*, which uses *basic* scale if the property values are absolute and *around_zero* scale otherwise.

Most of the algorithms scale on the basis of min and max bounds, hence future values may not fit that initial scale. The problem is mitigated if the normalized values are relative, which has also resulted in lower overfit and higher prediction accuracy in our experiments.

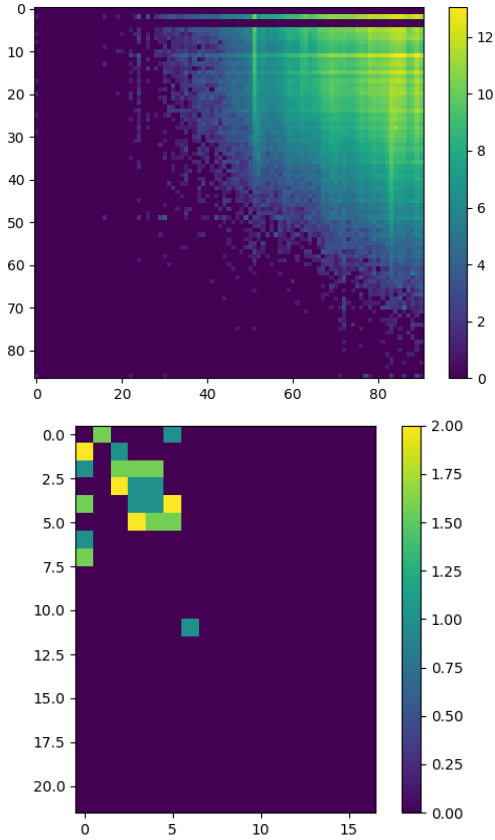

Figure 3: balanceLastSeenDistribution (up): we observe the difference in activity of the most active accounts (left) and richest accounts (down). The empty lines are seen because $\lfloor \log_{1.2}(x) \rfloor \notin \{1, 2, 4, 5\}$ for all $x$. User behavior and major market movements are more effectively seen in a time-lapse video of the distribution values (https://youtu.be/Dwwnxn1j6AQ). contractVolumeInERC20Distribution (down): we observe the different proportions of a crypto contract being used ($X$) and the amount of funds received from it ($Y$).

## 5.3 Dataset models

The dataset samples (input-output pairs) are created using a sliding window with a size of *wn* and step increments of 1 over a chosen set of normalized property values. Let us define a property's values as $prop_y$, where $y < propN$, a specific value in a property time series as $prop_{y,x}$, and $prop_t$ as the prediction target.

For $x \in [0, wn - 1]$, do:
    For $y \in [0, propN - 1]$, do:
        $win_{x,y} = prop_{y, x+step}$
$tar = prop_{t, wn+step}$

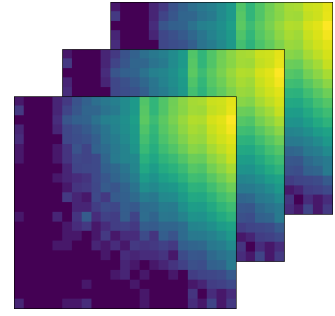

Figure 4: Visualized sample of the stacked layers model, where *wn* = 3 and the chosen properties consist of balanceLastSeenDistribution.

Where *win* is the window (unstructured dataset input), *tar* is the expected output, and *step* is incremented for each window.

In order to combine multiple normalized properties in a single dataset, we need the following models that define how their values are arranged in N-dimensional space.

*5.3.1 Matrix model.* Based on the LSTM network's 2D input shape, the matrix model defines one dataset sample as being the same as its corresponding *win*, without modifying its structure in any way.

Table 4 visualizes the structure of a dataset sample.

| Property | Value 0 | Value 1 | ... Value $wn - 1$ |
|---|---|---|---|
| Close value | 349$ | 358$ | |
| Number of TXs | 2876 | 1583 | |
| Operations with Contracts | 459 | 508 | |

Table 4: Example of a matrix model dataset sample.

*5.3.2 Stacked layers model.* A given set of property values (a column in *win*) has shape ($v_1, v_2, wn$), where $v_1 = 1$ and $v_2 = 1$ for all properties but the distributions (discussed in Sections 4.2 and 4.3); these properties with windowed values of 3D shape can't be modeled by a matrix. The stacked layers model is a 3D structure that matches input shape for the Convolutional Neural Network and allows modeling of 3D shaped property windows, preserving their spatial value.

The model defines a 3D matrix *mat* of shape ($values_1, values_2, wn$). A dataset sample *mat* is created as follows. For every set of windowed property values *vals*:

$$mat_{s_1+(0\ to\ v_1),\ s_2+(0\ to\ v_2),\ 0\ to\ wn} = vals_{0\ to\ v_1,\ 0\ to\ v_2,\ 0\ to\ wn}$$

Where $s_1$ and $s_2$ are set to the smallest integers where no values of other previously assigned properties are overridden. A visualized sample produced this model is seen in Figure 4.

## 6 GENERATED DATASETS

The experiment setup includes a number dataset configurations. The interval of the datasets is from 2017-03-01 to 2017-11-01. This range contains the largest fluctuations in Ethereum's history, the widespread media hoax about Ethereum (Roberts 2017), and the rise of ICOs (Catalini and Gans 2017).

Each complete dataset is split on *train* and *test* datasets. The first 7 months of the range build the *train* dataset, on the basis of which the neural network is trained. The last month is utilized by the *test* dataset, used for evaluation of the already trained network.

The dataset configurations are defined in Table 5. Window size (*wn*), prediction target and the normalization (*Norm*) are specified later in the experiments. As datasets 1 to 4 do not include blockchain data, they are used as a baseline for comparison.

| setN | Properties in dataset | Model |
|---|---|---|
| 1 | volumeFrom, volumeTo | matrix |
| 2 | volumeFrom_rel, volumeTo_rel | matrix |
| 3 | highPrice, volumeFrom, volumeTo | matrix |
| 4 | highPrice_rel, volumeFrom_rel, volumeTo_rel | matrix |
| 5 | accountBalanceDistribution | matrix |
| 6 | balanceLastSeenDistribution | stacked |
| 7 | contractBalanceLastSeenDistribution | stacked |
| 8 | balanceLastSeenDistribution, contractBalanceLastSeenDistribution, contractVolumeInERC20Distribution, accountBalanceDistribution | stacked |

Table 5: The dataset definitions.

## 7 NEURAL NETWORK ARCHITECTURES

The experimental setup includes a Long Short-Term Memory (LSTM) (Hochreiter and Schmidhuber 1997) and a Convolutional (CNN) (Lecun et al. 1998) models. Their network architectures are described in Figure 5.

The LSTM model, trained with datasets 1 to 4, is used as a baseline for comparison of the blockchain data approach.

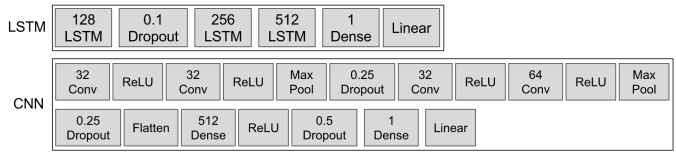

Figure 5: The architecture of the LSTM and Convoluitonal models respectively

## 8 EXPERIMENTS

We have performed hundreds of experiments with different dataset configurations and neural architectures. Out of those, the best ones are selected and grouped per prediction target.

The performance measurements are taken after inverse normalizing the network output and are therefore normalization independent. All measurements besides *sign* rely on the distance between the curves for prediction and actual values and are therefore directly incomparable between experiments with different targets.

The experimentally selected optimal training parameters include batch size of 16, $10^{-5}$ learning rate, and training via the Adam optimizer (Kingma and Ba 2014). All networks are trained with these parameters.

### 8.1 Predicting value (*highPrice*)

| N | setN | win | Norm | Netw | RMSE | $R^2$ | sign |
|---|---|---|---|---|---|---|---|
| 1 | 2 | 24 | image | LSTM | 90.4185 | 0.9131 | 0.5258 |
| 2 | 3 | 8 | prop | LSTM | 5.1060 | 0.9997 | 0.5095 |
| 3 | 5 | 24 | image | LSTM | 21.9306 | 0.9948 | 0.5054 |
| 4 | 6 | 104 | image | CNN | 16.2167 | 0.9972 | 0.5115 |
| 5 | 6 | 8 | image | CNN | 21.2845 | 0.9951 | 0.5033 |
| 6 | 7 | 104 | prop | CNN | 83.1030 | 0.9265 | 0.5290 |
| 7 | 7 | 24 | prop | CNN | 82.2199 | 0.9280 | 0.5367 |

Table 6: Results for target *highPrice*. The measurements are the best ones from all training epochs.

While the error scores on experiment 2 seem impressive, the network cheats by returning the inputted previous price data, instead of making an actual prediction. Experiment 1 does not input such information and forces the network to find an accurate solution. This can be seen in Figure 6. In order to avoid such scenarios, the input data in all CNN experiments omits previous prices.

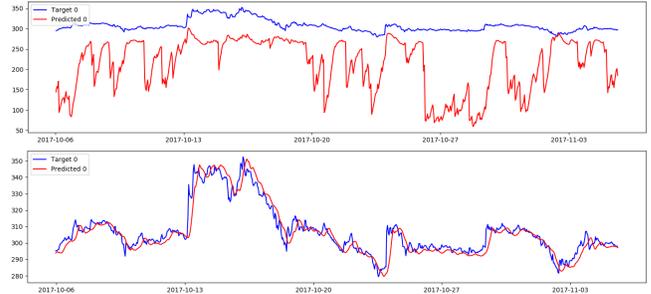

Figure 6: *highPrice*: The test results of exp 2 (bottom) and exp 1 (top). We observe how exp 2 achieves high accuracies by copying the previous price as a prediction and the result of omitting that data.

Exp 3 utilizes blockchain data and reaches an error reduction of 4 times compared to that of exp 1. Despite not using any market data, the network is able to predict multiple major market movements (Figure 7).

Exp 4 includes more valuable blockchain data and combined with CNN results in the lowest error scores for this target (26% lower than that of exp 3).

The high errors on experiments 6 and 7 may indicate that *prop* normalization is not suitable for distributions.

### 8.2 Predicting relative value (*highPrice_rel*)

Because the next relative change, as a value, is too different from the last relative change, a network cannot increase its score without predicting accurately. This explains the small error difference between exp 1 (volume data) and 2 (price & volume data).

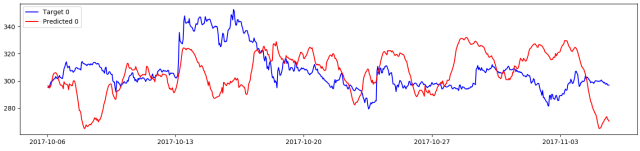

**Figure 7:** *highPrice*: **The test results of exp** 3. **On multiple occasions, we observe valid reactions to major market movements.**

| N | setN | win | Norm | Netw | RMSE | $R^2$ | sign |
|---|------|-----|------|------|------|-------|------|
| 1 | 1 | 104 | prop | CNN | 2.3652 | -0.0031 | 0.5502 |
| 2 | 4 | 104 | image | CNN | 2.3019 | 0.0498 | 0.4932 |
| 3 | 7 | 24 | image | CNN | 2.2825 | 0.0513 | 0.5340 |
| 4 | 8 | 8 | image | CNN | 2.2878 | 0.0461 | 0.5512 |
| 5 | 8 | 104 | prop | CNN | 9.1956 | -14.2060 | 0.5466 |

**Table 7: Results for target** *highPrice_rel*.

Experiment 4 utilizes 4 blockchain distributions and results in the lowest error and highest sign accuracy. However, when observing the prediction plot (Figure 8), the network is seen as less confident in its predictions. This may be mitigated with further network architecture and hyperparameter optimization.

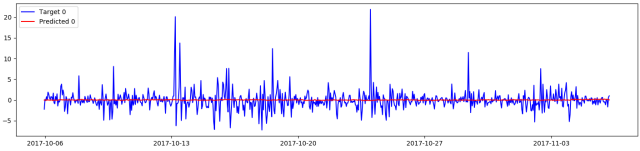

**Figure 8:** *highPrice_rel*: **The test results of exp** 4. **The small $\sigma$ of the predictions suggests lower prediction confidence.**

### 8.3 Predicting user growth (*uniqueAccounts_rel*)

| N | setN | win | Norm | Network | RMSE | $R^2$ |
|---|------|-----|------|---------|------|-------|
| 1 | 2 | 104 | prop | LSTM | 1139.972426 | 0.737456 |
| 2 | 1 | 104 | prop | LSTM | 718.971374 | 0.895567 |
| 3 | 7 | 8 | image | CNN | 812.303885 | 0.864637 |
| 4 | 6 | 24 | image | CNN | 639.336022 | 0.916289 |
| 5 | 8 | 8 | image | CNN | 622.716696 | 0.920449 |

**Table 8: Results for target** *uniqueAccounts_rel*.

This set of experiments aims to demonstrate how blockchain data can be used to predict other facts about the blockchain, like how the amount of unique users will grow.

Sign accuracy is irrelevant in this case, as the number of *uniqueAccounts* always increases.

When predicting blockchain facts, it is not a surprise that the experiments with market data (1 and 2) result in the lowest accuracies. Experiment 5 (Figure 9) yields impressive results in all measurements, which is likely due to the 4 distributions in its dataset.

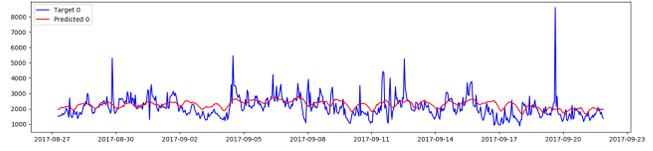

**Figure 9:** *uniqueAccounts_rel*: **The test results of exp** 5. **The prediction curve follows the actual values relatively accurately.**

## 9 RESULTS

Through the experiments, we have observed the problem from different angles, by training different networks, dataset configurations, and evaluation metrics.

The results have shown that Ethereum is too volatile to be predicted based solely on market data.

The best overall results are achieved using the proposed blockchain analysis approach, confirming our initial hypothesis that this kind of data can assist predictions. The approach has demonstrated effective 4 times reduction in error scores compared to only trade volume data. With the proposed Convolutional architecture, spatial account distributions, image normalization, and spatial dataset modeling, the error was additionally reduced by 26%. Training times were also reduced by a factor of 10.

Basic blockchain properties (as defined in Table 1) have demonstrated to be inefficient for the predictive task. These findings are consistent with the lower hourly prediction scores on the Bitcoin trading research by the Stanford research group (Madan 2014).

The spatial distributions have been consistently linked to the highest accuracy results, as they are able to analyze the blockchain raw data in more depth than other approaches.

The overall best dataset found combines all 4 of the spatial distributions (Number 8 in Table 5), modeled by the Stacked layers model (Section 5.3.2). The best found prediction target is relative high price, which forces the models to find the best possible predictive strategy.

The overall most accurate neural network is the proposed Convolutional model. Its ability to extract spatial meaning from the data likely contributes to its success in the distributions.

The best blockchain approach includes many parameters that have to be further explored before we can state the extent to which blockchain data can assist predictions.

## 10 CRYPTOCURRENCY PREDICTION FRAMEWORK

The technical implementation of the data processing, dataset generation and neural network evaluation processes is developed into an open source Cryptocurrency prediction framework. The framework allows for the prediction of a chosen cryptocurrency on the basis of user-defined property extraction and neural algorithms. The framework can be accessed at: https://github.com/Zvezdin/blockchain-predictor

The repository includes the necessary technical documentation for the reproduction or improvement of our experimental results.

All trained model weights, detailed training histories, and performance visualizations, are available here: https://goo.gl/uBQn4p.

## 11 FUTURE WORK

Throughout the research, we have evaluated not only a number of Deep Learning models but we have also built a reusable framework for data gathering, processing and storage of blockchain and market data. Once we had this framework, most of our actual research work was reduced to trying different network architectures and meta-parameters. While this process alone was challenging and demanding of both deep knowledge and creative thinking, we are excited by the possibilities to experiment with automating it.

Two recent publications by Google Research (Real et al. 2017) (Zoph and Le 2016) show that given enough computing power, we can create a "controller" algorithm which produces model designs for predicting the CIFAR10 dataset (Krizhevsky 2009), with performance on par with state-of-art models designed by humans. The process is called neural architecture search (NAS). The authors emphasize the enormous computational power required for achieving these results which was in order of $10^{20}$ computation. For comparison, most estimations about the computational power of the human brain are for between petascale ($10^{15}$) and exascale ($10^{18}$) operations per second (Martins et al. 2012).

We are interested in exploring NAS methodologies in order to find the most optimal neural configuration for blockchain data analysis.

Furthermore, we hope to develop a scalable decentralized architecture which allows remote nodes to provide computing power to "controller" algorithms for evaluating individual Deep Learning models. In such a system, Ethereum smart-contracts can be utilized to implement a decentralized marketplace for neural network model training and optimization.

Blockchain mining, which consumes enormous amounts of electrical power (Weinstein et al. 2018), might soon become obsolete due to the adoption of more efficient consensus algorithms like Proof-of-Stake. This presents an unique opportunity to utilize the abundant GPU mining resources for more reasonable work, like training Deep Learning models.

## 12 CONCLUSION

Blockchain technology enables a new class of tradable cryptotoken assets, thus creating new kinds of markets with numerous new use cases. Due to their decentralized and open nature, public blockchains provide an abundance of market-related data that has never been available before. Having every account balance and every transaction visible encourages us to seek a holistic view on the blockchain dynamics.

In our research project, we have evaluated different Deep Learning methods at their ability to find patterns in the Ethereum blockchain data and provide estimations about future measurements of the blockchain assets, including their market price. To facilitate our experiments, we have implemented a data gathering, processing, and storing framework that enabled us to deal with the enormous amount of data on the blockchain (more than 1000GB) and efficiently produce datasets for training.

We have tested both Recurrent and Convolutional models with different data sets and prediction targets. The most interesting results came from our Convolutional models which were working on account distribution histograms with one, two, and three dimensions. These histograms serve as "pictures" of the blockchain in a certain period of time and when we observe a time-lapse video of such 2D histograms (https://youtu.be/Dwwnxn1j6AQ), we can clearly see patterns that correlate to major market moves. After experimenting with these novel blockchain data representations, we found that Convolutional Deep Learning models are capable of providing estimations for the ETH token price based solely on such histogram data. Still, we believe that better results can be achieved by further optimizations of the model architectures and meta-parameters.

Since the search space for such optimizations is enormous, we will be focusing on creating efficient NAS approaches that automate the search process. We will also be investigating different ways in utilizing independent GPU resources in the common goal of finding a more efficient neural network.